\documentclass[11pt,twoside]{article}
\usepackage{asp2010}

\resetcounters

\begin{document}

\title{Low-mass stellar and substellar companions to sdB stars}
\author{S. Geier$^1$, L. Classen$^1$, P. Br\"unner$^1$, K. Nagel$^1$, V. Schaffenroth$^1$, C. Heuser$^1$, U. Heber$^1$, H. Drechsel$^1$, H. Edelmann$^1$, C. Koen$^2$, S. J. O'Toole$^3$, \\ L. Morales-Rueda$^4$
\affil{$^1$Dr. Karl Remeis-Observatory \& ECAP, Astronomical Institute,
Friedrich-Alexander University Erlangen-Nuremberg, Sternwartstr. 7, D-96049 Bamberg, Germany}
\affil{$^2$Department of Statistics, University of the Western Cape, Private Bag X17, Bellville 7535, South Africa}
\affil{$^3$Australian Astronomical Observatory, PO Box 296, Epping, NSW, 1710, Australia}
\affil{$^4$Department of Astrophysics, Faculty of Science, Radboud University Nijmegen, P.O. Box 9010, 6500 GL Nijmegen, NE}}

\begin{abstract}
It has been suggested that besides stellar companions, substellar objects in close orbits may be able to trigger mass loss in a common envelope phase and form hot subdwarfs. In an ongoing project we search for close substellar companions combining time resolved high resolution spectroscopy with photometry. We determine the fraction of as yet undetected radial velocity variable systems from a sample of 27 apparently single sdB stars to be $\simeq16\%$. We discovered low-mass stellar companions to the He-sdB CPD$-$20$^{\rm \circ}$\,1123 and the pulsator KPD\,0629$-$0016. The brown dwarf reported to orbit the eclipsing binary SDSS\,J0820+0008 could be confirmed by an analysis of high resolution spectra taken with UVES. Reflection effects have been detected in the light curves of the known sdB binaries CPD\,$-$64$^{\rm \circ}$\,481 and BPS\,CS\,22169$-$0001. The inclinations of these systems must be much higher than expected and the most likely companion masses are in the substellar regime. Finally, we determined the orbit of the sdB binary PHL\,457, which has a very small radial velocity amplitude and may host the lowest mass substellar companion known. The implications of these new results for the open question of sdB formation are discussed.
\end{abstract}

\section{Introduction}

About half of the sdB stars reside in close binaries with periods ranging from a few hours to a few days \citep{maxted01,napiwotzki04}. Because the components' separation in these systems is much less than the size of the subdwarf progenitor during its red-giant phase, these systems must have experienced a common-envelope and spiral-in phase \citep{han02,han03}. 

Although the common-envelope ejection channel is not yet properly understood in detail, it provides a reasonable explanation for the strong mass loss required to form sdB stars. However, for about half of all known subdwarfs there is no evidence for close stellar companions as no radial velocity variations are found. Among other formation scenarios, the merger of two helium white dwarfs has often been suggested to explain the origin of single sdB stars \citep{han02,han03}. 

\citet{soker98} suggested that substellar objects like brown dwarfs and planets may also be swallowed by their host star and that common envelope ejection could form hot subdwarfs. Substellar objects with masses higher than $\simeq10\,M_{\rm J}$ were predicted to survive the common envelope phase and end up in a close orbit around the stellar remnant, while planets with lower masses would entirely evaporate or merge with the stellar core. The stellar remnant is predicted to lose most of its envelope and evolve towards the extreme horizontal branch (EHB). A similar scenario has been proposed to explain the formation of apparently single low mass white dwarfs \citep{nelemans98}. The discovery of a brown dwarf with a mass of $0.053\pm0.006\,M_{\rm \odot}$ in an $0.08\,{\rm d}$ orbit around such a white dwarf supports this scenario and shows that substellar companions can influence the outcome of stellar evolution \citep{maxted06}.

The planet discovered to orbit the sdB pulsator V\,391\,Peg with a period of $1\,170\,{\rm d}$ and a separation of $1.7\,{\rm AU}$ was the first planet found to have survived the red-giant phase of its host star \citep{silvotti07}. Serendipitous discoveries of two substellar companions around the eclipsing sdB binary HW\,Vir \citep{lee09} and one brown dwarf around the similar system HS\,0705$+$6700 \citep{qian09} followed. These substellar companions to hot subdwarfs have rather wide orbits, were not engulfed by the red giant progenitors and therefore could not have influenced the evolution of their host stars. But the fact that substellar companions in wide orbits around sdBs seem to be common suggests that similar objects closer to their host stars might exist as well \citep[for a review see][]{schuh10}. 

Here we present new results from our ongoing search for sdBs with close substellar companions using high-resolution, time-resolved spectroscopy as well as time-resolved photometry. First we determine the fraction of as yet undetected RV variable systems from a sample of apparently single sdB stars and review the known candidate systems. Then we present newly discovered systems with low-mass stellar as well as substellar companions and discuss the implications of these systems for sdB formation.

\section{Are all 'single' sdBs RV variable?}\label{sec2}

The most important drawback of the radial velocity (RV) method is the unknown inclination of the binaries. From an RV curve of a single-lined binary alone only a lower limit for the companion mass can be derived. For a single object it is therefore impossible to prove the existence of a substellar companion, because a more massive stellar companion seen at low inclination cannot be excluded. Most sdB binaries are single-lined systems and only lower limits can be put on the masses of their companions usually assuming a canonical EHB mass of $\simeq0.47\,M_{\rm \odot}$. Since about $50\%$ of all known sdBs are in close binary systems with stellar companions, there must be a certain number of such systems seen at low inclinations. Furthermore, most RV variable systems have been identified from medium resolution time-resolved spectroscopy \citep[e.g.][]{copperwheat11} and the fraction of binaries with  variabilities too small to be detected at such resolutions is therefore not well constrained. In principle, all apparently single sdBs could have close, but yet undetected companions.

A large sample of sdBs has to be studied to decide whether the fraction of systems with small RV variations is consistent with the low-inclination extension of the known sdB binary population or not. A higher fraction than expected would be an indication for a population of substellar companions. Up to now our sample consists of 27 bright single-lined sdB stars. We used high resolution spectra ($R=48\,000$) obtained with ESO-2.2m/FEROS. Each star has been observed several times and the timespans between the observations range from days to years. We chose a set of sharp, unblended metal lines with accurate rest wavelengths, fitted Gaussian and Lorentzian profiles to the metal lines and determined the RVs. The errors ranged from $\simeq0.3$ to $2.0\,{\rm km\,s^{-1}}$. To check the wavelength calibration for systematic errors we used telluric features as well as nightsky emission lines. 

Four stars were found to show significant RV variability. The RV shifts range from $0.5$ to $12.7\,{\rm km\,s^{-1}}$. Follow-up photometry and high resolution spectroscopy are necessary to exclude pulsational variability and derive the orbital parameters of these binaries. Constraints can then be put on the companion masses. No significant RV variations were found in the  rest of the sample. We deduce that any undetected RV variation of a programme star has to be lower than its RV measurement uncertainty. \citet{soker98} suggests that these objects should be more massive than $10\,M_{\rm J}$, otherwise they would have been destroyed during the CE phase, and should have orbital periods of the order of $\simeq10\,{\rm d}$. 

Adopting this period we derive an upper limit for the binary mass function, which translates into an upper limit for $M_{\rm 2}\sin{i}$ if we assume the canonical mass of $0.47\,M_{\rm \odot}$ for the sdB. In this way tight constraints can be put on possible substellar companions. Substellar companions with $M_{\rm 2}\sin{i}=5-35\,M_{\rm J}$ can be excluded. In conclusion $\simeq\,16\%$ of the apparently single stars in our sample show RV variations. The true fraction is expected to be higher, because the accuracy is limited by the S/N of the spectra in most cases. Substellar companions in close orbits can be most likely excluded in $\simeq84\%$ of our sample. 

Our RV survey showed that most of the apparently single sdB stars do indeed not show RV variations of more than $\simeq1.0\,{\rm km\,s^{-1}}$ on timescales from days to years. Nevertheless, $\simeq\,16\%$ do show small RV variations. However, it has still to be shown whether these variations are due to orbital motion.

\section{Candidate sdB systems with substellar companions}

\subsection{Candidates from literature}

Several close binary systems with possible substellar companions have been found. A companion in the planetary mass range was reported to orbit the bright sdB \textbf{HD\,149382} based on an RV curve with very small amplitude ($P=2.39\,{\rm d}$,$K=2.3\,{\rm km\,s^{-1}}$). However, \citet{jacobs11} and \citet{norris11} did not detect significant RV variability in this period range and excluded the presence of a planetary companion. We obtained high resolution follow-up spectra with AAT/CYCLOPS ($R=80\,000$). Consistent with the results of \citet{jacobs11} and \citet{norris11} we were not able to verify the RV variability on timescales of a few days reported in \citet{geier09}.

The eclipsing binary \textbf{AA\,Dor} ($P\simeq0.26\,{\rm d}$, $K=40\,{\rm km\,s^{-1}}$) has a companion with a mass close to the hydrogen burning limit (Rauch et al. these proceedings). \textbf{PG\,1329+159} is a reflection effect binary with very similar orbital parameters ($P\simeq0.25\,{\rm d}$, $K=40\,{\rm km\,s^{-1}}$) and a minimum companion mass ($0.08\,M_{\rm \odot}$) right at the border region between stars and brown dwarfs \citep{morales03,geier10}. 

The very short-period reflection effect binary \textbf{PG\,1017$-$086} ($P\simeq0.073\,{\rm d}$, $K=51\,{\rm km\,s^{-1}}$) may have a brown dwarf companion \citep[$>0.05\,M_{\rm \odot}$,][]{maxted02,geier10} as well. The very similar but eclipsing system \textbf{J1622+4730} ($P\simeq0.07\,{\rm d}$, $K=47\,{\rm km\,s^{-1}}$) has been found in the course of the MUCHFUSS project (Geier et al. these proceedings). 

The reflection effect binary \textbf{KBS\,13} (KIC\,1868650) most likely has a substellar companion as well. \citet{for08} determined not only the orbital parameters of this binary ($P\simeq0.29\,{\rm d}$, $K=23\,{\rm km\,s^{-1}}$), but also the projected rotational velocity and surface gravity of the sdB ($v_{\rm rot}\sin{i}=23.3\pm1.1\,{\rm km\,s^{-1}}$, $\log{g}=5.87$). Assuming the canonical mass for the sdB as well as orbital synchronisation of the sdB, the mass of the companion can be calculated as described in \citet{geier10} to be $0.055\,M_{\rm \odot}$. Because the rotational velocity of sdBs in binaries with such low-mass companions has been found to be slower than predicted for a synchronised orbit (see Sect.~3.3, 3.4, 3.5), the companion mass derived for KBS\,13 must be regarded as an upper limit, while a lower limit is provided by the binary mass function ($0.046\,M_{\rm \odot}$). The companion mass of KBS\,13 is therefore tightly constrained. The extremely accurate Kepler light curve of this star can be used to verify this conclusion \citep{oestensen10}.

\subsection{Low-mass stellar companions to CPD$-$20$^{\rm \circ}$\,1123 and KPD\,0629$-$0016}

\textbf{CPD\,$-$20$^{\rm \circ}$\,1123 (Albus 1)} was discovered to be a bright He-sdB by \citet{vennes07}. Except for the unique double-lined spectroscopic binary PG\,1544+488 \citep{ahmad04} none of the known He-sdBs is known to reside in a close binary system. We obtained medium resolution time resolved spectroscopy ($R=3400$) of this star using ESO-NTT/EMMI and found it to be RV variable. Follow-up spectroscopy was taken with FEROS and an orbital solution could be determined ($P=2.3\,{\rm d}$, $K=44.3\,{\rm km\,s^{-1}}$). The minimum mass of the unseen companion is $0.21\,M_{\rm \odot}$. Whether the companion is a compact object like a white dwarf or a low-mass MS star is unclear. Due to the rather long period a reflection effect indicative of a cool MS star would not be detectable. CPD\,$-$20$^{\rm \circ}$\,1123 is the first He-sdB with a close unseen companion. 

\textbf{KPD\,0629$-$0016} is the only known pulsating sdB star (type sdBV$_{\rm s}$) in the CoRoT field \citep[][and references therein]{charpinet10}. We detected RV variability of the order of $K\simeq60\,{\rm km\,s^{-1}}$ based on medium resolution spectra taken with ESO-NTT/EMMI and ESO-NTT/EFOSC2, which are too high to be caused by pulsations. The orbital period could not yet be determined from available data, but is most likely shorter than one day, and the minimum mass of the unseen companion is of the order of $0.2\,M_{\rm \odot}$. More observations are needed to obtain a unique orbital solution and clarify the nature of the companion. The high-precision CoRoT light curve may be of great help in this respect.

\begin{figure}
\begin{center}
  \includegraphics[width=8cm]{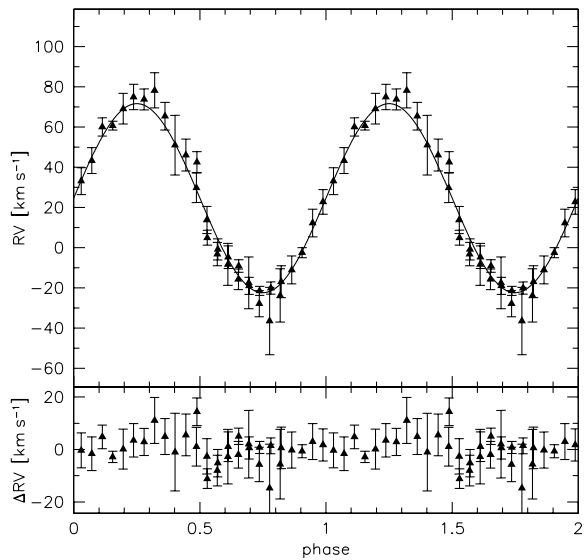}
\end{center}
\caption{Phased RV curve of SDSS\,J0820+0008. The RVs have been measured from high resolution spectra taken with UVES.}
\label{fig:hd149}
\end{figure}

\subsection{SDSS\,J0820+0008 - Brown dwarf confirmed}

The eclipsing sdB binary SDSS\,J0820+0008 was reported to host a brown dwarf companion \citep{geier11d}. In order to verify these results and to constrain the binary parameters better we obtained $3.5\,{\rm hr}$ of consecutive time resolved high resolution spectroscopy with ESO-VLT/UVES ($R\simeq40\,000$) covering the whole orbit of the system. No spectral features from the irradiated companion could be detected. The orbital parameters derived from the new dataset ($P=0.097\,{\rm d}$, $K=47.0\pm1.2\,{\rm km\,s^{-1}}$ see Fig.~\ref{fig:hd149}) are in perfect agreement with the previous results from \citet{geier11d}. Atmospheric parameters of the sdB primary are determined for the first time with high accuracy ($T_{\rm eff}=25900\,{\rm K}$, $\log{g}=5.42$) from spectra taken during the total eclipse of the companion. Comparing the surface gravity derived from the spectroscopic analysis with the value derived from the light curve solution the mass of both binary components can be constrained. With the new results we can exclude an sdB mass of more than $\simeq0.5\,M_{\rm \odot}$ and  a companion mass exceeding $\simeq0.07\,M_{\rm \odot}$ confirming the substellar nature of the companion. 

Using the UVES spectra we were able to measure the projected rotational velocity of the sdB for the first time with high accuracy ($v_{\rm rot}\sin{i}=67\pm2\,{\rm km\,s^{-1}}$). In contrast to our expectations and despite the very short period of the system \citep{geier10} this velocity turned out to be much too low for the sdB to rotate synchronously with its orbital motion ($v_{\rm rot}\sin{i}_{\rm syncro}\simeq120\,{\rm km\,s^{-1}}$). Another interesting result is a significant shift of the system velocity $\Delta \gamma=+15\pm1\,{\rm km\,s^{-1}}$ with respect to the results of \citet{geier11d}. A systematic zero-point shift of this order is rather unlikely. Another explanation would be the presence of a third unseen component in the system. More observations are needed to investigate this further.

\subsection{CPD\,$-$64$^{\rm \circ}$\,481 - Reflection effect reveals brown dwarf companion}

\begin{figure}
\begin{center}
  \includegraphics[width=8cm]{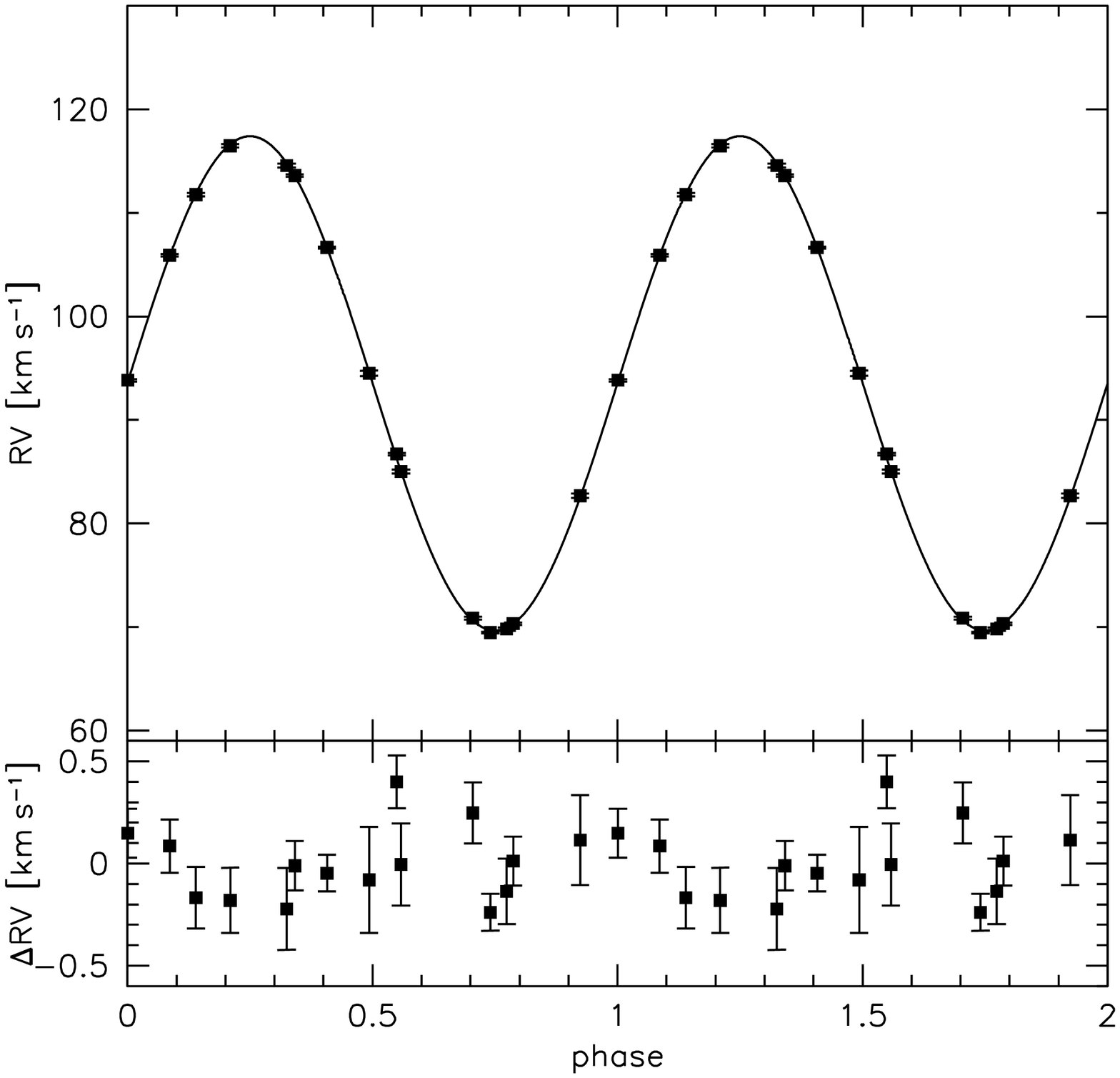}
  \includegraphics[angle=-90,width=10cm]{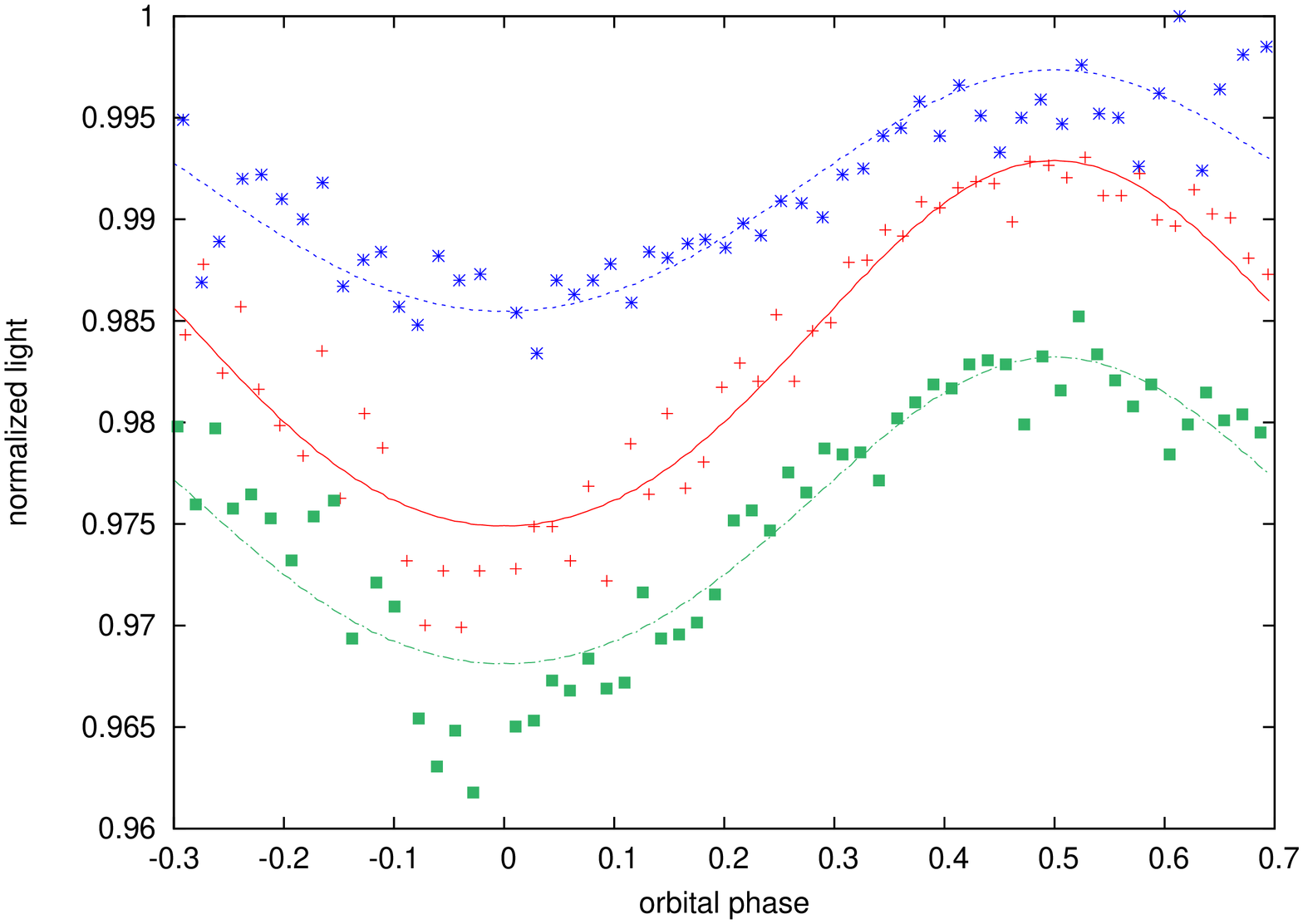}
\end{center}
\caption{Upper panel: Phased Radial velocity curve of CPD\,$-$64$^{\rm \circ}$\,481. Lower panel: Light curves of CPD\,$-$64$^{\rm \circ}$\,481 taken at SAAO (B,R,V-band) phased to the orbital period. The reflection effect is clearly visible as well as the absence of eclipses.}
\label{fig:cpd}
\end{figure}

\citet{edelmann05} discovered CPD\,$-$64$^{\rm \circ}$\,481 to be a close sdB binary with very small RV amplitude and concluded that the companion may be a substellar object. We measured the RVs of CPD\,$-$64$^{\rm \circ}$\,481 from FEROS spectra as described in Sect.~\ref{sec2} and determined the orbital parameters ($P=0.277263\pm0.000005\,{\rm d}$, $K=23.90\pm0.05\,{\rm km\,s^{-1}}$ see Fig.~\ref{fig:cpd} upper panel), which are perfectly consistent with the parameters derived by \citet{edelmann05} using the same data. \citet{geier10} measured a very low $v_{\rm rot}\sin{i}$ of the subdwarf primary and derived a much higher companion mass in the stellar regime assuming tidally locked rotation. In this scenario, the binary would be seen nearly pole-on ($i\simeq7^{\rm \circ}$).

However, photometric follow-up revealed that this scenario must be incorrect. We took a multi-colour light curve with the the SAAO STE4 CCD camera mounted on the SAAO 1.0-m telescope and detected a sinusoidal variation with the orbital period characteristic for a reflection effect (see Fig.~\ref{fig:cpd} lower panel). This modulation is caused by the changing light contribution of the irradiated companion when orbiting around the hot primary. Such a reflection effect is only detectable if the binary has a high inclination, which means that the low inclination determined in \citet{geier10} must be highly underestimated, and that the sdB rotates much slower than synchronisation. 

The inclination of the binary will be constrained by a light curve analysis \citep[e.g.][]{geier11d}. Preliminary solutions favour inclinations between $60^{\rm \circ}$ and $70^{\rm \circ}$. Adopting the canonical sdB mass of $0.47\,M_{\rm \odot}$ the mass of the unseen companion most likely ranges between $0.050\,M_{\rm \odot}$ and $0.055\,M_{\rm \odot}$, well below the stellar mass limit. The detection of a reflection effect in the light curve of CPD\,$-$64$^{\rm \circ}$\,481 therefore revealed that the companion of CPD\,$-$64$^{\rm \circ}$\,481 is most likely a brown dwarf and that the rotation of the sdB cannot be tidally locked.

\subsection{BPS\,CS\,22169$-$0001 - Small reflection effect and non-synchronised rotation}

BPS\,CS\,22169$-$0001 was discovered to be a close sdB binary with very small RV amplitude by \citet{edelmann05}. The conclusion that the companion may be  substellar was again questioned by \citet{geier10}, who measured a low $v_{\rm rot}\sin{i}$, assumed tidal synchronisation of the sdB and constrained the companion mass to be stellar. Again, a very small reflection effect ($0.1\%$) was detected in a light curve of this binary obtained at SAAO, which allowed us to constrain the orbital period better ($P\simeq0.214\,{\rm d}$). The new period is somewhat longer than the one given in \citet{edelmann05} which was based on just a few FEROS spectra. Adopting this period and using the nine FEROS spectra as well as four spectra obtained with the Coude spectrograph mounted at the 2.7m McDonald telescope we derive an RV-semiamplitude $K=16.2\,{\rm km\,s^{-1}}$. 

Dropping the assumption of orbital synchronisation the minimum mass of the unseen companion is as low as $0.026\,M_{\rm \odot}$, but due to the very small amplitude of the light curve variation the inclination angle is expected to be smaller than in the case of CPD\,$-$64$^{\rm \circ}$\,481. Anyway, the inclination would have to be smaller than $\simeq25^{\rm \circ}$ to lift the companion above the stellar mass limit. A combined spectroscopic and photometric analysis is necessary to constrain the companion mass more tightly. 

\subsection{PHL\,457 - The lowest mass substellar companion?}

\begin{figure}
\begin{center}
  \includegraphics[width=8cm]{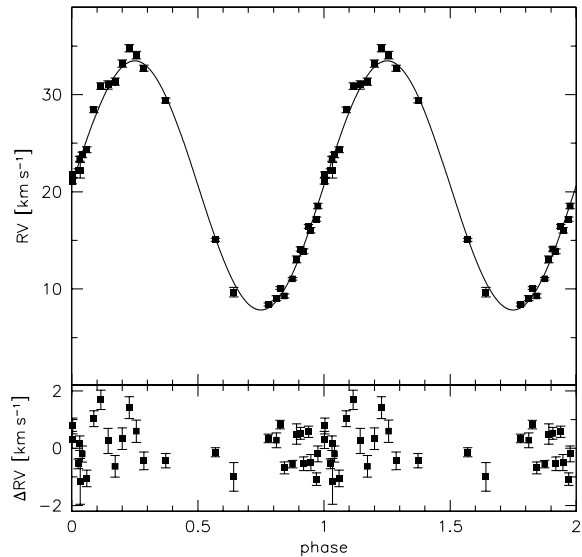}
\end{center}
\caption{Phased Radial velocity curve of PHL\,457.}
\label{fig:phl}
\end{figure}

PHL\,457 was discovered to be RV variable by \citet{edelmann05}, but the data were not sufficient to determine the orbital parameters. We obtained 26 additional spectra with FEROS and derived the orbital solution with high statistical significance ($P=0.3128\pm0.0007\,{\rm d}$, $K=12.80\pm0.08\,{\rm km\,s^{-1}}$ see Fig.~\ref{fig:phl}). The minimum mass of the unseen companion is $0.026\,M_{\rm \odot}$ and therefore very well below the stellar mass limit. Furthermore, \citet{blanchette08} found PHL\,457 to be a long period sdB pulsator (sdBV$_{\rm s}$) with a possible reflection effect present in the light curve. The latter is important because it hints at a quite high inclination, which means that the actual companion mass cannot be much higher than the lower limit. However, the data are not sufficient to clarify this issue (Green priv. comm.). 

\section{Conclusion}

\begin{figure}
\begin{center}
  \includegraphics[width=7cm]{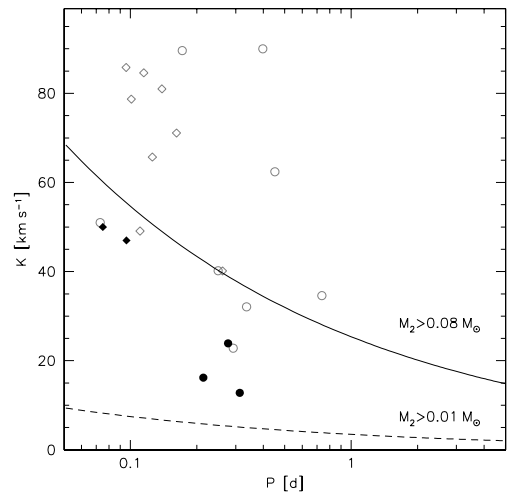}
\end{center}
\caption{The RV semiamplitudes of all known reflection effect sdB binaries plotted against their orbital periods. Eclipsing systems are plotted as diamonds, non-eclipsing systems as circles. Open symbols mark all known systems from the literature \citep[see][and references therein]{geier11b}, while the filled symbols mark the newly discovered systems presented here and in Geier et al. (these proceedings). The solid line marks the border between the stellar and the substellar mass regime, the dashed line the border between brown dwarfs and planets.}
\label{fig:substellar}
\end{figure}

In Fig.~\ref{fig:substellar} the orbital parameters of all known reflection effect sdB binaries are plotted. From the 21 systems known, 11 may have substellar companions. Since only about $100$ close sdB binaries with orbital solutions are known, the fraction of systems with low-mass companions exceeds $20\%$. The fraction of possible substellar companions is higher than $10\%$. Due to selection effects, the true fraction must be higher. Reflection effects or eclipses necessary to clarify the nature of the companions are more likely to be found in close and highly inclined binaries. Recent discoveries of low-mass objects around sdBs with periods of more than ten days \citep{geier11c,barlow11} indicate that such systems do exist as well and that they might be quite numerous.

It can be seen in Fig.~\ref{fig:substellar} that the minimum companion mass seems to decrease with longer orbital periods. This is contrary to what is expected from common envelope evolution theory: The smaller the mass of the companion, the more the system has to shrink before enough energy and angular momentum is transferred to the common envelope to be ejected. However, this scenario assumes that the companion survives the CE phase. But the fate of substellar objects engulfed by close red giants is unclear. \citet{soker98} argues that substellar objects might also evaporate during the CE phase or merge with the core of the red giant \citep[see also][]{politano08}. Most recently, we discovered the rapidly rotating single sdB EC\,22081$-$1916, the possible outcome of such a common envelope merger event \citep[][Geier et al. these proceedings]{geier11a}. The destruction of the companions during or shortly after the CE phase might explain the lack of lowest-mass objects with very short periods seen in Fig.~\ref{fig:substellar}. Those objects might have come too close to the red-giant core to survive.

We conclude that close substellar companions to sdB stars do exist and that they are by no means rare. They play a significant role in the formation of sdB stars. They may either form close binaries and survive, merge with the red-giant core creating rapidly rotating single sdBs or evaporate in the red-giant envelope. In the latter case they create ordinary single sdBs that rotate slowly (see also Geier et al. these proceedings). 

\end{document}